\documentclass[]{spie}  %>>> use for US letter paper
\usepackage{amsmath,amsfonts,amssymb}
\usepackage{graphicx}
\usepackage[colorlinks=true, allcolors=blue]{hyperref}
\usepackage[normalem]{ulem}

\title{Illinois Express Quantum Network (IEQNET): Metropolitan-scale experimental quantum networking over deployed optical fiber}

\author[1,2]{Joaquin Chung}
\author[3,4]{Gregory Kanter}
\author[5,6]{Nikolai Lauk}
\author[5,6]{Raju Valivarthi}
\author[7]{Wenji Wu}
\author[7,8,9]{Russell R. Ceballos}
\author[5,6,7]{Cristián Peña}
\author[5,6,10]{Neil Sinclair}
\author[4]{Jordan Thomas}
\author[5,6]{Si Xie}
\author[1,2,11]{Rajkumar Kettimuthu}
\author[4,12]{Prem Kumar}
\author[7]{Panagiotis Spentzouris}
\author[5,6]{Maria Spiropulu}

\affil[1]{Data Science and Learning Division, Argonne National Laboratory, 9700 S Cass Ave. Lemont, IL. 60439, USA}
\affil[2]{The University of Chicago Consortium for Advanced Science and Engineering, 924 E 57th St, Chicago, IL 60637, USA}
\affil[3]{NuCrypt LLC, 1840 Oak Ave, Evanston, IL 60201, USA}
\affil[4]{Center for Photonic Communication and Computing, Department of Electrical and Computer Engineering, McCormick School of Engineering and Applied Science, Northwestern University, 2145 Sheridan Road, Evanston, IL 60208-3118, USA}
\affil[5]{Division of Physics, Mathematics and Astronomy, California Institute of Technology, 1200 E. California Blvd., Pasadena, CA 91125, USA }
\affil[6]{Alliance for Quantum Technologies (AQT), California Institute of Technology, 1200 E. California Blvd., Pasadena, CA 91125, USA}
\affil[7]{Fermi National Accelerator Laboratory, Kirk Rd and Pine St, Batavia, IL 60510, USA}
\affil[8]{Department of Chemistry, Physics, and Engineering Studies, Chicago State University, 9501 S. King Dr., Chicago, IL 60628, USA}
\affil[9]{External Member Pittsburgh Quantum Institute, B4 Thaw Hall, 4061 O'Hara Street, Pittsburgh, PA 15260, USA}
\affil[10]{John A. Paulson School of Engineering and Applied Sciences, Harvard University, 29 Oxford St., Cambridge, MA 02138, USA}
\affil[11]{The Northwestern Argonne Institute of Science and Engineering (NAISE), 2205 Tech Drive Suite 1-160, Evanston, IL 60208, USA}
\affil[12]{Department of Physics and Astronomy, Northwestern University, 2145 Sheridan Road, Evanston, IL 60208-3112, USA}

\begin{document} 
\maketitle

\begin{abstract}
The Illinois Express Quantum Network (IEQNET) is a program to realize metro-scale quantum networking over deployed optical fiber using currently available technology. 
IEQNET consists of multiple sites that are geographically dispersed in the Chicago metropolitan area. 
Each site has one or more quantum nodes (Q-nodes) representing the communication parties in a quantum network. 
Q-nodes generate or measure quantum signals such as entangled photons and communicate the results via standard, classical, means.
The entangled photons in IEQNET nodes are generated at multiple wavelengths, and are selectively distributed to the desired users via optical switches. 
Here we describe the network architecture of IEQNET, including the Internet-inspired layered hierarchy that leverages software-defined-networking (SDN) technology to perform traditional wavelength routing and assignment between the Q-nodes. 
Specifically, SDN decouples the control and data planes, with the control plane being entirely classical. 
Issues associated with synchronization, calibration, network monitoring, and scheduling will be discussed. 
An important goal of IEQNET is demonstrating the extent to which the control plane can coexist with the data plane using the same fiber lines. 
This goal is furthered by the use of tunable narrow-band optical filtering at the receivers and, at least in some cases, a wide wavelength separation between the quantum and classical channels. 
We envision IEQNET to aid in developing robust and practical quantum networks by demonstrating metro-scale quantum communication tasks such as entanglement distribution and quantum-state teleportation.  
\end{abstract}

%\keywords{}

\section{INTRODUCTION}
\label{sec:intro} 
Analogous to classical optical networks, it is expected that quantum optical networks will be used to interconnect quantum processors including quantum computers and sensors. While it is attractive to leverage as much of the traditional fiber-optic infrastructure as possible, quantum optical signals have different characteristics as compared to conventional optical data.
For instance, they are sensitive to loss and even minuscule amounts of unintended light can manifest itself as detrimental noise. 
Moreover, quantum signals are incompatible with often-used optical amplifiers and end-point electrical terminations. As such, we expect quantum networks to require custom engineering while also benefiting from as much compatibility with traditional networks as is practical.

The Illinois Express Quantum Network (IEQNET) is a program for realizing a metropolitan-scale quantum network over deployed fiber optics cables. 
While new technologies, such as quantum repeaters, will be needed to realize the long-term goal of quantum networking over long distances, IEQNET focuses on leveraging only the currently available technology (with allowance for future upgrades as technology develops). 
Quantum communication networks often target the quantum key distribution (QKD) application, which in many implementations, such as the BB84 protocol, does not require the distribution of entanglement. 
On the other hand, to target advanced applications beyond QKD, such as networked quantum computing and quantum sensing, distributing entanglement under network control will be required. Therefore, IEQNET is designed to provide quantum entanglement distribution and information transfer by means of quantum teleportation as a core network service. 
Moreover, IEQNET aims to perform these tasks while simultaneously allowing (strong) classical signals to share the same fiber, both for the purposes of enabling communications to support quantum applications and for independent co-existence of high-rate data channels. Although challenging due to the aforementioned differing properties of quantum and classical signals, this type of network potentially offers the greatest utility, largest impact and fastest approach to widespread deployment. 
In this paper we introduce our plan to realize these functions over a metro-scale network, focusing on entanglement distribution for simplicity, and discuss some of the major issues that need to be addressed to enable reliable network operation. 
While prior works have demonstrated quantum teleportation between two locations using in-ground fiber \citenum{Valivarthi2016, Sun2017} and other entanglement distribution networks, e.g. \citenum{Joshi2020}, there is a pressing need to develop the necessary system architecture to advance entanglement-based technologies into practical usage.
This includes achieving better integration into a classical network framework, scaling to more users, reaching longer link distances, allowing co-existing data channels, and automating the hardware and software control of network operations.

\section{IEQNET's design and architecture}
\begin{figure}[htbp]
\centering
  \includegraphics[width=0.80\textwidth]{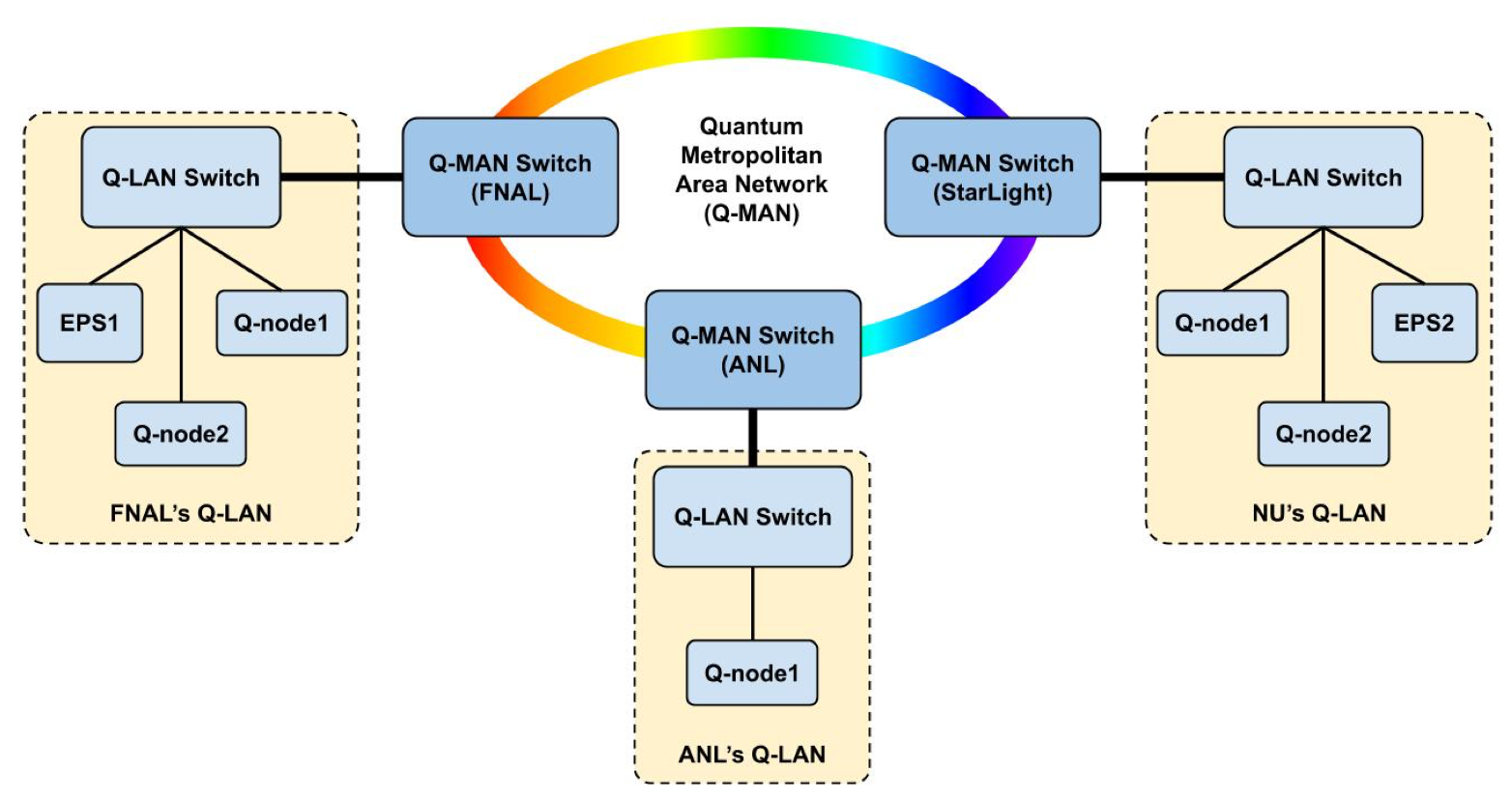}
\caption{IEQNET's topology. The proposed network consists of three logically independent quantum local area networks that are located at Fermilab (FNAL), Argonne National Lab (ANL) and Northwestern University (NU) campuses. The three Q-LANs are connecting to each other forming a meshed network topology.}\label{Fig:IEQNET}
\end{figure}

As illustrated in Fig. \ref{Fig:IEQNET}, IEQNET consists of multiple sites that are geographically dispersed in the Chicago metropolitan area. 
The sites are physically located at Northwestern University, Fermi National Accelerator Laboratory (FNAL), Argonne National Laboratory, and a Chicago-based data center (Starlight). 
Three logically independent quantum local area networks (Q-LANs), namely Fermilab's Q-LAN, Argonne Q-LAN, and Northwestern University Q-LAN are contained within IEQNET.
Each Q-LAN has one or more quantum nodes (Q-nodes), which can communicate data as well as generate and/or measure quantum signals. 
Q-nodes will be connected to software-defined networking (SDN)-enabled optical switches by way of optical fibers.
The optical switches are connected by a dedicated classical communication channel (ESnet ChiExpress MAN) and additional dark (i.e., no co-existing classical signals) fibers between FNAL and StarLight for quantum signals.
This approach yields a meshed network topology.

\subsection{Quantum nodes}
\begin{figure}[htbp]
\centering
  \includegraphics[width=0.80\textwidth]{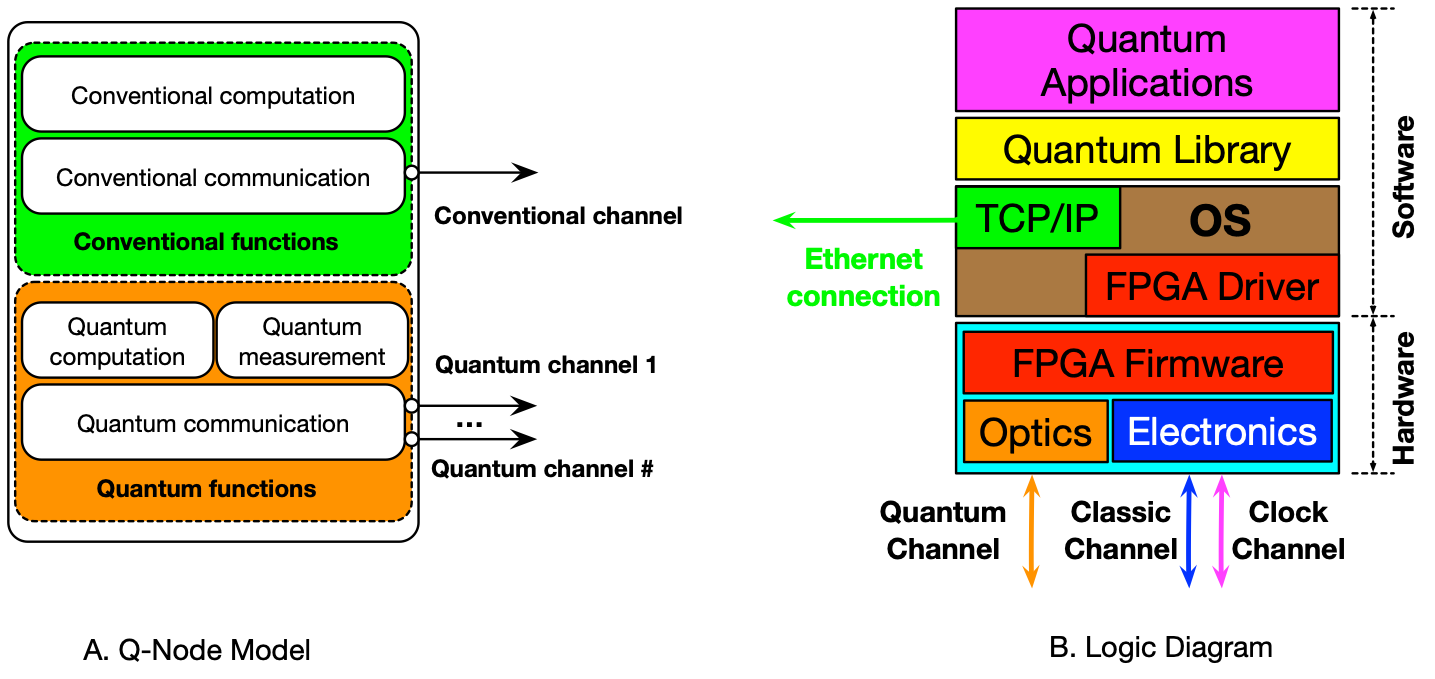}
\caption{IEQNET's quantum node model A. together with node's logic diagram B. Each node performs conventional and quantum functions and has optical and electronic interfaces that are controlled by a FPGA with additional software layers running on top.}\label{Fig:Qnodes}
\end{figure}

Q-nodes in IEQNET, depicted in Fig. \ref{Fig:Qnodes}, are much like their classical counterparts (i.e., nodes) the communication parties in quantum networks.
A Q-node performs both conventional and quantum functions.
The quantum functions that the node has to execute depend on the node type.
These range from single and entangled photonic qubit generation to qubit measurements and processing, including Bell-state measurements.
A Q-node also performs conventional functions such as classical computation and communication. 
For example, a Q-node will exchange messages with other Q-nodes via conventional traffic channels to exchange the results of quantum measurements, e.g., to determine correlations, or carry out quantum protocols. 
A Q-node is assigned an Internet Protocol (IP) address to uniquely identify the node. 
Meanwhile the endpoints of each conventional channel is identified and addressed by a physical address. 
In addition, a Q-node enumerates and numbers the quantum channels it connects to.
The \textit{\textless Q-node ip, quantum channel \#\textgreater} pair is used to identify and address a particular quantum channel endpoint.

Conventional channels serve three purposes in IEQNET: (i) as communication channels between Q-nodes to exchange data and timing messages in order to enable quantum (correlation) measurements, (ii) embedded signals to allow calibration and stabilization functions (e.g. polarization stabilization, calibration of quantum basis measurements, or fiber-delay stabilization), (iii) as test channels for experimentation, e.g. to determine the limits of classical/quantum channel coexistence. 
We note that the calibration/stabilization could use the quantum signal directly, but the use of larger magnitude classical signals when possible can provide enhanced functionality, such as reducing set-up time when the network changes the user connections or allows high-rate feedback control to compensate time-varying parameters (like polarization).

Routing is a fundamental network function. 
Multi-hop networks require a means of selecting paths through the network. 
Due to technology immaturity in quantum memory and quantum computation, IEQNET does not perform routing in the quantum domain, such as entanglement routing. 
Instead, SDN technology is used to perform traditional \textit{wavelength routing and assignment} in optical networks to establish quantum paths between Q-nodes, or between Q-nodes and entangled photon pair sources (EPSs), as appropriate. 
This is largely implemented by controlling the SDN-enabled optical switch (SOS) and the tunable band-pass filter (TBF) at the optical input of a node (see Sec. \ref{sec:ControlPlane} for more details).

\subsection{Entanglement generation}
\begin{figure}[htbp]
\centering
  \includegraphics[width=0.80\textwidth]{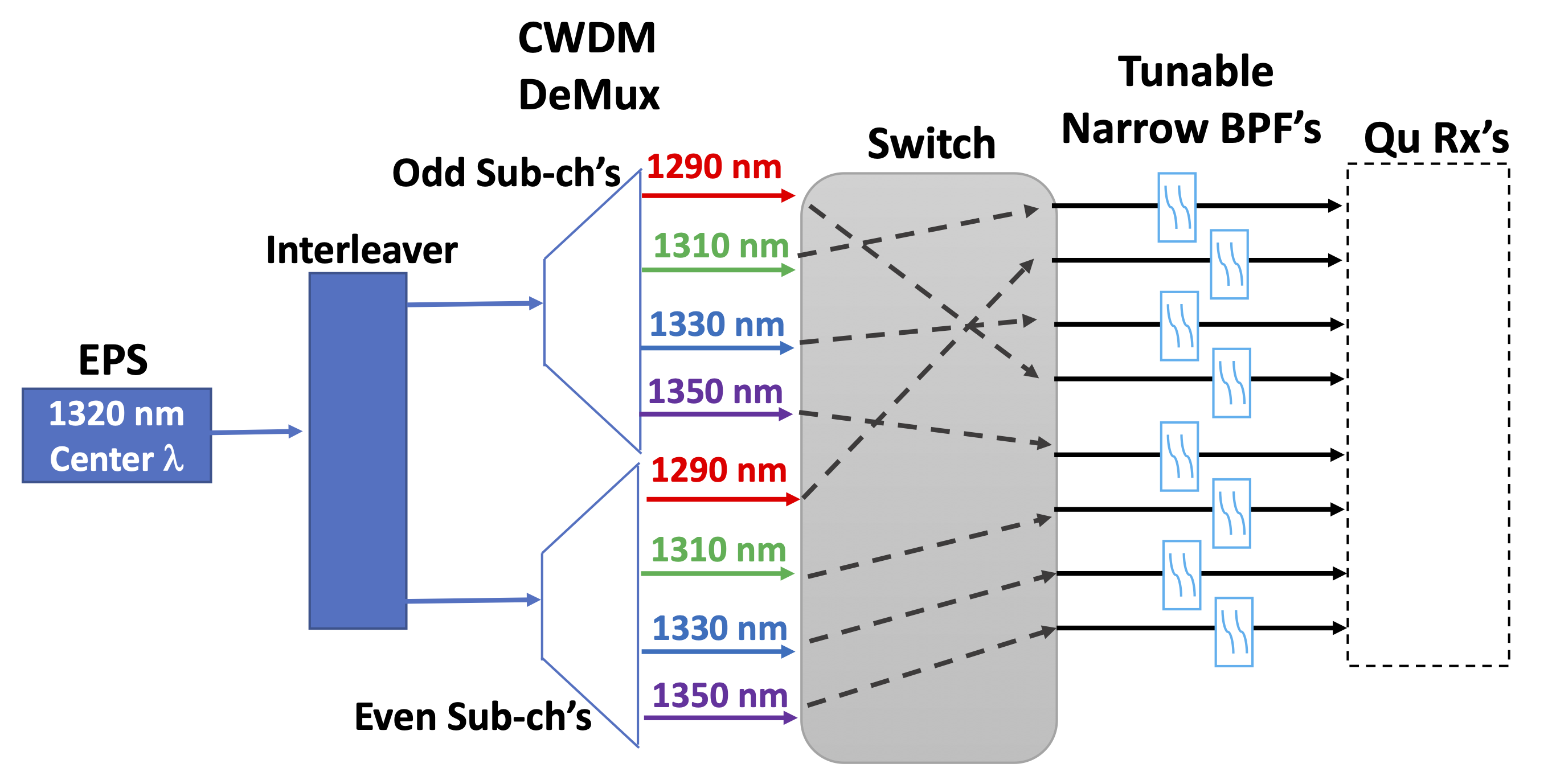}
\caption{Multiplexed entangled photon pair source design for IEQNET. A broadband nonlinear optical crystal generates N/2 entangled photon pairs at N wavelengths. Coarse wavelength division multiplexer (CWDM) is used to spatially separate the pairs in N individual fibers. The signals are routed to the receivers by means of optical switches and are isolated at the receiver's end using tunable bandpass filters (BPF). }\label{Fig:EPS}
\end{figure}

To establish quantum communication between a pair of users in the network, we use EPSs to generate bipartite quantum signals at N wavelengths (e.g. N=8). 
These signals are then distributed throughout IEQNET, allowing a maximum of N/2 pairs of users to share an entangled state, a prerequisite for performing any task.
The EPS can be realized in several ways \citenum{Srivathsan2013, Orieux2017, Caspani2017}, one common approach is to use engineered $\chi^{(2)}$ optical nonlinearities in optical fiber-coupled solid-state waveguides \citenum{Arahira2013}. 
This process generates entangled photon pairs over a broad bandwidth, for instance ~6 THz \citenum{Arahira2013}, which means the number of available channels is largely limited by the availability and specification of low-loss optical filters (see Fig. \ref{Fig:EPS}). 
A commonly used International Telecommunication Union-defined frequency channel-spacing for conventional communication is 100 GHz (i.e. dense wavelength division multiplexing, DWDM). 
With this approach, about N = 64 routable quantum channels would be achievable, with wavelengths in the original or conventional band (i.e. O- or C-band), at 1.3 $\mu$m and 1.5 $\mu$m, respectively. 
The photon pairs could be entangled in their time-of-arrival or polarization properties for instance, generating time-bin and polarization entanglement, respectively, depending on the EPS used. 
Broadly speaking, polarization entanglement does not require stabilized interferometers to perform measurements, and thus provides some advantages, however it degrades in the presence of fiber birefringence and requires polarization stabilization systems.

As indicated in Fig. \ref{Fig:EPS} N EPS wavelengths are spatially separated into N fibers and feed an M$\times$M optical switch, where M \textgreater\ N (e.g. M=32). 
Such optical switches greatly expand the number of possible users without adding much insertion loss (e.g. 1 dB). Insertion loss is particularly critical for quantum networks as the quantum signals cannot be amplified. Elements (such as the optical switch) that are common to both the signal and idler photons attenuate both wavelengths, e.g. 3 dB additional insertion loss leads to a reduction of coincidence count rates by a factor of 4. 
The architecture allows for a central multi-wavelength EPS to service many users, and also allows for the possibility of quantum-channel redundancy for added resiliency. Multiple SDN-enabled optical switches (SOS) can be interconnected to reach a very large number of potential users with one or a small number of centralized multi-wavelength entangled sources. 
The number of simultaneous users is limited by the number of wavelength channels that can be matched by the spectrum of each entangled photon source.

Moreover, unlike a typical lossy (e.g. 5-8 dB) wavelength-selective switch (WSS), the SOS operates over a wide optical bandwidth allowing for both C-band (1525-1565 nm) and O-band (1290-1350 nm) operation.
This is important because some EPS stations will operate in the O-band in order to facilitate co-existence with classical data. 
Ensuring the entangled photon pair wavelengths are at an optical frequency that is \textgreater\ 20 THz higher than the classical data greatly reduces spontaneous Raman noise leaking into the quantum channel from the data channels \citenum{Chapuran2009}. 
This allows higher co-existing classical power levels and thus higher net data rates. 

The wide wavelength difference between quantum and classical channels also allows the bands to be combined or separated with very low insertion loss, a function that may have to be performed multiple times, for example if lossy or noisy elements like WSS’s or fiber amplifiers (which are already deployed in existing conventional networks) need to be bypassed by the quantum channel.
Another strategy for allowing classical/quantum co-existence is matched filtering in the quantum domain, which cuts off unnecessary wide-band noise. 
For pulses of the desired pulse duration, which for our purposes is \textgreater\ 50 ps for minimal timing constraints when interfering pulses in a Bell state measurement (BSM) as part of a teleportation protocol \citenum{Valivarthi2020}, matched filtering relies on narrow optical filters (\textless\ 10 GHz) just before the receiver. 
In cases where the incoming photon wavelength is chosen by the network, a tunable band-pass filter (TBF) at the Q-node serves the function of both narrowing the bandwidth and selecting the channel to be received. 

\subsection{Layered Architecture}
Inspired by the layered architecture of the classical Internet, IEQNET implements a similar layered quantum networking architecture, which describes how quantum network functions are vertically composed to provide increasingly complex capabilities, see Fig. \ref{Fig:Layers}.
Other research groups have proposed similar layered architectures for quantum networks.
For instance, Dahlberg et al.~\citenum{dahlberg2019} (QuTech at Delft University, Netherlands) proposed quantum network stack that perfectly maps the names of the TCP/IP stack (i.e., they have Application, Transport, Network, Link, and Physical layers).
Similarly, Alshowkan et al.~\citenum{alshowkan2021} (Oak Ridge National Laboratory, USA) proposed an architecture that considers Application, Transport, Link, and Physical layers.
The authors purposely exclude a network layer because they consider this is only needed for connecting multiple independent networks like in the case of the classical Internet.
IEQNET layered quantum networking architecture relies on four key vertical layers that we describe in detail in the following subsections.

\subsubsection{Quantum Physical Layer}
The quantum Physical Layer deals with the physical connectivity of two communicating quantum nodes, including all intermediate optical devices and fibers.
It defines quantum channel frequencies, signal rates, photon pulses used to represent quantum signals, etc.
Both QuTech and Oak Ridge National Laboratory (ORNL) architectures consider a Physical layer that encompasses all physical components of the quantum networks.

\subsubsection{Quantum Link Layer} 
IEQNET's quantum Link layer resides in Q-nodes and handles the transmission of quantum signals and messages across quantum channels.
It is also responsible for monitoring quantum link status. 
On the contrary, QuTech's Link layer provides robust entanglement services between Q-nodes in the same Q-LAN.
ORNL provides even another disagreeing definition.
For them, the Link layer partitions the available spectral bandwidth and routes it between endpoints.

\subsubsection{Quantum Network Layer} 
The quantum Network Layer refers to using SDN to perform traditional \textit{wavelength routing and assignment} in optical networks to establish quantum paths between Q-nodes.
QuTech considers the Network layer as a provider of long-distance entanglement, which is outside the scope of IEQNET.
As mentioned before, ORNL considers the Network layer for Q-LANs and they have pushed the routing functionality to their Link layer.

\subsubsection{Quantum Service Layer}
IEQNET's quantum Service Layer refers to providing quantum entanglement distribution service to users and applications. 
Quantum service layer functions include EPS management, such as entanglement generation and entanglement distribution.
Our service layer is equivalent to ORNL's Transport layer as they also consider entanglement distribution and end-to-end service.
On the contrary, QuTech pushed the ``local'' entanglement distribution down to their Link layer.

\subsubsection{Quantum Application Layer}
We consider quantum applications those that can make use of end-to-end entanglement distributed by IEQNET (e.g., quantum teleportation and quantum sensing).
In this regard, ORNL's definition of the Application layer aligns with ours.
However, QuTech considers quantum teleportation a service of the Transport layer and therefore considers applications those that make use of teleportation.

\begin{figure}[htbp]
\centering
  \includegraphics[width=0.6\textwidth]{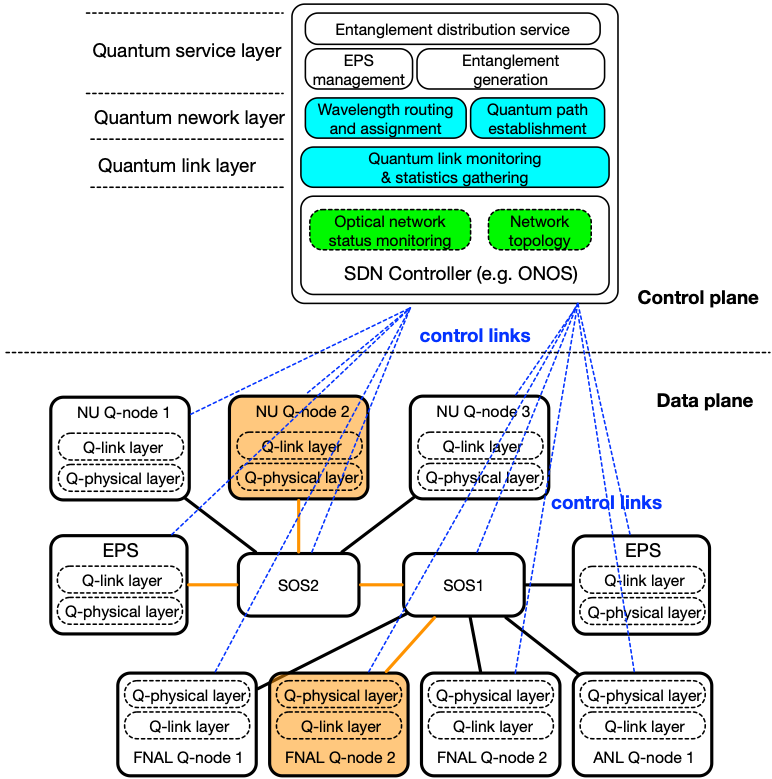}
\caption{IEQNET's layered architecture with centralized control. See main text for details.}\label{Fig:Layers}
\end{figure}

\subsection{Control plane design}\label{sec:ControlPlane}
IEQNET uses a centralized control approach. As illustrated in Fig. \ref{Fig:Layers}, SDN decouples the control and data plane. 
In the control plane, the SDN controller monitors the status of IEQNET (e.g., loss on fiber links, status of optical switches, etc.). 
IEQNET control and management software runs on top of the controller to perform various functions, such as wavelength routing and assignment for quantum channels, dynamical quantum path establishment, and quantum entanglement distribution.
In the data plane, quantum signals and messages are transmitted across quantum channels between Q-nodes, or between Q-nodes and EPSs.

To perform wavelength routing and assignment (RWA), the SDN controller first discovers the network topology by communicating via classical communication protocols (e.g., OpenFlow) with optical switches. The network topology is considered as a unidirected graph $G$, where each node in the graph represents a node in the network, and each link in the graph represents a link in the network. To find the best available path in the network we consider several metrics (e.g., fiber loss, insertion loss, polarization dependent loss, etc.) as the weight for each link and apply path finding algorithms for routing.
RWA is typically formulated as a multicommodity flow problem. This is an NP-hard problem, typically solved with heuristic algorithms which are suitable for and SDN approach.

\section{Entanglement distribution protocol}
As an example for the quantum network use cases we here discuss the entanglement distribution case. The corresponding IEQNET protocol is characterized by the following message flow between the entities:
\begin{enumerate}
    \item A user sends a request to IEQNET controller for entanglement distribution between Q-node-1 and Q-node-2.
    \begin{itemize}
        \item Qubit types (polarization, time-bin,...)
        \item Start/END time
        \item Q-node pair
        \item Common measurement basis for calibration
    \end{itemize}
    \item IEQNET controller analyzes the request and chooses an EPS that meets the requirement, or the request is rejected.
    \item IEQNET controller executes path routing \& wavelength assignment and establishes paths among involved entities.

    \item IEQNET notifies Q-node-1 and Q-node-2 that paths are established
    \item Q-node-1, Q-node-2, and EPS test and verify paths. If there are failures, go back to step 3. Otherwise, send acknowledgement (ACKs) to IEQNET controller.
    \begin{itemize}
        \item Verification of the path from EPS to Q-node1 is done in two stages. First a classical light (some light of known power for e.g.) is sent from EPS to Q-node1. The received power is measured at Q-node-1 and communicated to the IEQNET controller. This will be used to establish losses.
        \item Next step is to send quantum light from EPS which is measured in the single photon detectors (SPDs) and upon receiving expected clicks, the Q-node-1 will send a message to the IEQNET controller saying that the path has been verified. Similar steps can be followed to verify the path between EPS and Q-node-2. Assuming the EPS is calibrated and the losses from EPS to Q-nodes are also known from the previous step, the expected clicks will be $R*\eta$ where $R$ is the rate of photon generation at EPS and $\eta$ the loss between the EPS and the Q-node. Measure the “noise” on the detector by switching off the quantum light from EPS. Calculate the ratio of noise/rates and if the ratio is less than some specified threshold (e.g. $1/6$) declare it as a success. 
Similar steps are followed for the clock paths too.
    \end{itemize}
    \item IEQNET performs calibration and preparation for entanglement distribution.
    \begin{itemize}
        \item The classical signals to be exchanged between the two receiver stations for sync and channel monitoring are assigned the wavelength of $\lambda_{\textrm{sync}}=$ 1551.72 nm (DWDM ITU Channel C32). These signals can go back and forth between the two receiver stations.
        \item Measurement basis alignment
        \begin{enumerate}
            \item Polarization Qubit: 
            \begin{itemize}
                \item Add classic alignment light at EPS \citenum{Wang2013}.
                \item IEQNET controller notifies EPS  to send out alignment signals. The controller will specify the basis.
                \item Q-node tunes its analyzer to maximize the output so that measurement basis(es) are aligned with the EPS basis(es).
                \item Q-node stops and notifies IEQNET.
            \end{itemize}
            \item Time Bin qubit:
            \begin{itemize}
                \item IEQNET controller notifies EPS to send early,early photons. Q-node will identify the “early” timing with respect to the local clocks. Then the EPS will send late, late photons which will then allow Q-nodes to identify the “late” timing.
                \item If the receivers are using the same detectors to detect in time and phase basis then IEQNET controller asks EPS to send the entangled photons and receiver stations identify the so-called early, middle and late bins from the histograms of the detection signals obtained locally.
                \item For interferometer alignment at both locations we add classic interferometer alignment light at EPS.
                \item IEQNET controller notifies EPS  to send out interferometer alignment signals. The controller will specify the phase.
                \item Q-node tunes its interferometer phase to maximize the output to the corresponding phase so that phases are aligned at both the Q-nodes.
                \item Q-node stops and notifies IEQNET.
            \end{itemize}
        \end{enumerate}
        \item In the calibrate phase, the entangled photons received by both are detected in a common measurement basis which is agreed upon before. The detection results are encoded onto a laser with wavelength ($\lambda_{\textrm{sync}}$) and exchanged back and forth between the two stations.
        \item These results are used then to achieve bit level sync using local time-taggers / coincidence analyzers at each receiving station.
        \item This is also where the correlation delay can be established. Ideally the controller has a range of possible delay values that may work for measuring correlations at a given node, which it communicates to the node. While the measurement bases are aligned the node can scan the electrical delay (in integer clock units) until heightened correlations are observed. Typically the node can determine if the delay setting is likely to be right or wrong after ~10 coincidence counts are measured (standard deviation of the measurement is then $\sqrt{10}$). If it is likely to be correct then another 10 coincidence counts should verify it.
        \item Proceed to entangle phase.
        \item Note that $\lambda_{\textrm{quantum}}$, $\lambda_{\textrm{clk}}$, and $\lambda_{\textrm{sync}}$ can all be multiplexed onto a single fiber strand or just the classical ones onto a different fiber strand as envisioned for the initial phases of the project.
        \item Choose a suitable duty cycle of the experiment and repeat the calibration step at the start of each cycle.
    \end{itemize}
    \item EPS, Q-node-1, Q-node-2, and other related entities notify READY to IEQNET controller.
    \item IEQNET controller sends START to EPS.
    \item EPS starts to distribute entanglement pairs; Q-node-1 and Q-node-2 receive entangled photons.
    \item Q-node-1 and Q-node-2 measures the entangled photons, and exchange results. They store the results locally until they have long enough e-bit string for the required application and communicate END to the IEQNET controller.
    \item IEQNET controller stops EPS with an END command.
    \item The results will be posted to the IEQNET controller and will be stored.
\end{enumerate}
The teleportation protocol can be devised in similar manner. 
\section{Conclusions}
We presented details for a proposed metropolitan-scale, repeater-less quantum network. 
We discussed the network topology and defined quantum nodes and their functions, such as the quantum receiver or entanglement source node.
Further, drawing from the (classical) Internet we outlined the layered network architecture that uses SDN-based technology to decouple data and control plane and enables centralized control.
Lastly, we discussed an exemplary entanglement distribution protocol as one of the possible network use cases.

We envision IEQNET to aid in developing free-running,  practical, quantum network controls and demonstrating metro-scale quantum operations including entanglement distribution and eventually quantum teleportation. 

\acknowledgments % equivalent to \section*{ACKNOWLEDGMENTS}       
 IEQNET is funded by the Department of Energy's Advanced Scientific Computing Research Transparent Optical Quantum Networks for Distributed Science program, but no government endorsement is implied.

\bibliography{SPIEbib} % bibliography data in report.bib
\bibliographystyle{spiebib} % makes bibtex use spiebib.bst

\end{document}